\documentclass[aps,pre,amsfonts,showpacs]{revtex4} 
\usepackage{epsfig,amsopn}
\usepackage{graphicx}

\begin{document}
\def\be{\begin{equation}}
\def\ee{\end{equation}}
\def\bea{\begin{eqnarray}}
\def\eea{\end{eqnarray}}
\def\bml{\begin{mathletters}}
\def\eml{\end{mathletters}}
\def\l{\label}
\def\tnd{\rightarrow}
\def\eqn#1{(~\ref{eq:#1}~)}
\def\av#1{{\langle  #1 \rangle}}


\title{A simple sandpile model of active-absorbing state transitions}

\author{Kavita Jain}
\affiliation{Institut 
f\"ur Theoretische Physik, Universit\"at zu K\"oln, 50937 K\"oln, Germany}

\begin{abstract}
We study a simple sandpile model of active-absorbing state transitions  
in which a particle can hop out of a site only if the number of particles 
at that site is above a certain threshold. We 
show that the active phase has product measure whereas nontrivial 
correlations are found numerically in the absorbing phase. It is argued that 
the system relaxes to the latter phase slower than exponentially. The 
critical behavior of this model is found to be different from that of the 
other known universality classes. 
\end{abstract}

\pacs{02.50.-r, 64.60.-i, 05.65.+b} 

\maketitle


\section{Introduction}
\label{intro}

Active-absorbing state transitions form an important class of nonequilibrium
phase transitions whose best studied example is directed 
percolation (DP) \cite{abs}. A critical behavior different 
from DP is possible in the presence 
of additional symmetries as in parity-conserving DP \cite{pcdp}. 
Sandpile models in which the activity occurs 
when the local density exceeds a certain threshold
can also exhibit such transitions, provided the total 
density (``energy'') is conserved \cite{fesmrev}. 
Depending on the nature of the coupling between the activity and the local 
density, the conservation law may \cite{fesm1} or may not \cite{dd}
change the critical behavior from DP. Further, the non-DP behavior can 
belong to different universality classes such as 
C-DP which describes stochastic sandpiles with linear 
coupling \cite{cdp1,cdp2}.  

The models mentioned above have resisted an exact solution so far and 
have been studied extensively using numerical simulations and field-theoretic 
techniques. In this article, we introduce and study the steady state and the 
dynamics of 
a simple FESM for which some exact results can be found. 
As we shall see, the critical behavior 
of this model is different from that of the other known classes of FESMs.

In Section~\ref{ss}, we define the model and study 
its steady state. 
In our model, only \emph{one} particle hops out of a site to a nearest
neighbor if the number of particles at that site exceed a certain threshold. 
In the high-density active phase, this model defines a zero range 
process (ZRP) whose 
steady state is known to be of product measure form in all 
dimensions \cite{spitzer,evans}. In the low-density absorbing phase, the 
steady state is not unique and depends on the initial conditions. 
We determine the critical behavior of this model by studying an order 
parameter and a correlation function close to the critical density. 
The time-dependent properties of this model are described in 
Section~\ref{dyn}. 
We argue that the relaxation 
dynamics are the same as that of a two-species annihilation process; in 
particular, we find a stretched exponential decay in the inactive 
phase, which has been seen numerically in other 
sandpiles also \cite{sexpo}. Our arguments are supported by Monte Carlo 
simulations. 


\section{The model and its steady state}
\label{ss}

The model is defined on a $d$-dimensional hypercubic lattice of length
$L$ with periodic boundary conditions. Each site can hold an arbitrary number
of particles with unit mass. We call a site active if the mass at this site 
is greater than $m_{thr}$ otherwise inactive. A particle hops out of 
an active site to a nearest neighbor with equal probability. In 
continuous time, the hopping rate $u(m)$ at a site having $m$ particles is 
given by
\be
u(m)=\cases { 1/2d & {, $m > m_{thr}$}  \cr
                0  & {, $m \leq m_{thr} \;\;.$}
} \l{rates}
\ee
The dynamics conserve the total density $\rho=M/V$ where $M$ is the total
number of particles in the system and $V \equiv L^{d}$ is the volume 
of the system. In one dimension with $m_{thr}=1$, this model can be mapped 
to a conserved lattice gas (CLG) model by regarding the sites as holes and 
mass as 
particle clusters. In the CLG language, a particle with one
occupied neighbor can only hop to an empty nearest neighbor whereas the
isolated particles do not move \cite{clg}. 

To see the phase transition, we define all the particles at a site to be 
immobile if the mass $m \leq m_{thr}$ at this site. For $m > m_{thr}$, the 
first $m_{thr}$ particles are said to be immobile and the 
rest $m-m_{thr}$ mobile. If the initial number of particles is less than 
$m_{thr} V$, a mobile particle diffuses around till it reaches a site 
with $m < m_{thr}$ whereupon it becomes immobile. Thus, the number of mobile 
particles eventually becomes zero. On the other hand, if the total number of 
particles in the initial state exceed $m_{thr} V$, then the number 
of mobile particles decrease till it reaches $M-m_{thr} V$.
Thus, there is a phase transition 
at the critical density $\rho_{c}=m_{thr}$ from an inactive phase 
with only immobile particles 
to an active one with a finite number of mobile particles, as the total 
density is increased. 
For sake of simplicity, we choose $m_{thr}=1$ in the following discussion. 


\subsection{The active phase}
\l{active}

We first note that the probability $P_{k}(0,t)$ of having $m=0$ at 
site $k$ at time $t$ is a monotonically decreasing function in time since
\be
\frac{\partial P_{k}(0,t)}{\partial t}=-\frac{1}{2d} \sum_{\delta}
\sum_{m=2} P_{k,k+\delta}(0,m,t) \;\;,
\ee 
where $P_{k,k+\delta}(m',m,t)$ is the joint probability that the site $k$
and its neighbor $k+\delta$ have mass $m'$ and $m$ respectively at time $t$. 
In the active phase, since $\rho > 1$, it follows that the probability 
$P(0)$ of having an empty site is zero. It is then easy to check that 
the condition of detailed balance holds with the 
steady state distribution $P(C \equiv \{ m_{1},...,m_{V} \})$ being 
equally likely. Since the number of 
ways in which $M-V$ particles can be partitioned in $V$ cells is given 
by $Z={M-1 \choose V-1}$ \cite{huang}, we have $P(C)=1/Z$. Using a similar 
reasoning, the
mass distribution $P(m)={M-m-1 \choose V-2}/Z$ can be obtained. Thus, the 
activity $S=\sum_{m=2} P(m)=(M-V)/(M-1)$. Near the 
critical point, $S \sim (\rho-\rho_{c})^{\beta}$ with $\beta=1$ in all 
dimensions. Intuitively, as $\rho \tnd \rho+1/V$, due to the absence of 
empty sites, the newly added particle typically becomes a mobile 
particle, leading to a linear growth of activity. This argument fails for 
the models studied in \cite{fesm1} in which more than one particle can leave 
an active site thus creating empty sites. In such cases, one may expect 
the activity to grow nonlinearly.

Since $P(0)=0$ for our model, the exponent $\beta$ is expected to be unity 
even if $u(m)$ 
is mass-dependent for $m > m_{thr}$. 
This can be shown by realising that the active phase is a special case of 
ZRP for which $P(C) \sim \prod_{k=1}^{V} \; f(m_{k})$ 
where the marginal $f(m_{k})$ is given by 
\be
f(m_{k})=\cases{\prod_{n_{k}=2}^{m_{k}} 1/u(n_{k}) 
& {, $ \;\; m_{k} \ge 2$} \cr
             1 & {, $\;\; m_{k}=0, 1 \;\;.$}
} \l{fs}
\ee
The distribution $P(m)=v^{m} f(m)/Z$ is obtained by using the measure 
$P(C)$ above with fixed density constraint. Here the normalisation 
constant $Z= \sum_{m=1} v^{m} f(m)$ and the fugacity 
$v$ is determined by the conserved particle number condition which can 
be written as
\be
\rho-1= \frac{1}{Z}\sum_{m=2} (m-1) v^{m} f(m) \;\;.
\ee
Close to the critical density, the fugacity $v \rightarrow 0$ and we 
can approximate $\rho-1$ by the first term in the sum above. Using this in 
the expression 
for the activity $S=\sum_{m=2} v^{m} f(m)/Z$ for 
$v \rightarrow 0$, it follows that $\beta=1$. 

Typically,  the density-density correlation function  
$C(r,\rho)=\langle m_{0} \; m_{r} \rangle -\rho^{2}$ decays exponentially with 
$r$ and can be used to define a static correlation length 
$\xi \sim (\rho-\rho_{c})^{- \nu_{\perp}}$. Since this phase has 
product measure, the correlation function $C(r,\rho)$ is zero 
in the thermodynamic limit for all $r$. This implies that 
the correlation length $\xi_{>}=0$ and 
the exponent $\nu_{\perp}^{>}$ is undefined in this phase.  


\subsection{The inactive phase and the critical point}

For $\rho < \rho_{c}$, the system is in the inactive phase and the number of 
absorbing states is given by ${V \choose M}$. Unlike the active phase, here 
the steady state is neither unique nor has product measure even if the 
initial state has. To see this, 
let us prepare an initial state in which a site is occupied by a monomer with 
probability $p_{1}$, a dimer with 
probability $p_{2}$ and remains empty with $p_{0}=1-p_{1}-p_{2}$. 
Then, for $p_{1}=0$, the final state will have one-clusters 
with only even number of ones whereas for $p_{1} \neq 0$, the 
configurations with odd number of ones are also allowed. Thus different 
initial conditions give different solutions for $P(C)$ so that the steady 
state is not unique. Further, this is clearly not a product measure state for  
which the weight of all configurations at a fixed density is same. 

Using Monte Carlo simulations in one dimension, we measured the correlation 
function   
$C(r,\rho)=\langle m_{0} \; m_{r} \rangle -\rho^{2}$ 
where $m$ is either $0$ or 
$1$. Since the steady state depends on the initial condition, the angular 
brackets denote spatial averaging for a given initial condition. We initially 
distribute particles independently at each site 
with the mass chosen from a Poisson distribution. 
As shown in Fig.~\ref{ic}, the behavior of the correlation function 
$C(r,\rho)$ when averaged over such initial conditions is similar to when only 
spatial averaging is done for a fixed initial configuration.
Therefore, in the following discussion, 
we will carry out ensemble 
averaging as well. 

The data in the inset of Fig.~\ref{xi} for $C(r,\rho)$ at various 
densities for a large, one-dimensional system supports the claim that the 
product measure does not hold in this phase. For densities away from unity, 
the correlation function decays fast and as one approaches the critical 
point, the decay gets slower. The magnitude of $C(r,\rho)$ at small $r$ 
drops for $\rho \tnd 1$ since at the critical density, we must obtain 
$C(r,\rho_{c})=0$. Also, since the total mass is conserved, we have 
\be
\av{M m_{0}}= L \rho^{2}=\av{m_{0}^2}+\av{m_{0} m_{1}}+...+\av{m_{0} m_{L-1}} \;\;.
\ee
Thus, the correlation function obeys the sum rule $\sum_{r} C(r,\rho)=0$ due 
to which $C(r,\rho)$ is not positive for all $r$. 
Finally, as shown in 
Fig.~\ref{xi}, a data collapse for $C(r, \rho)$ for densities close to the 
critical density can be obtained if we assume
\be
C(r,\rho) \approx \xi_{<}^{-1} g(r/ \xi_{<})\;\;,\;\; 
\xi_{<}=(1- \rho)^{-2} \;\;. \l{n0nr}
\ee
The above scaling form is consistent with vanishing $C(r,\rho)$ at 
the critical point. Thus, the static correlation length 
$\xi_{<} \sim (\rho_{c}-\rho)^{-\nu_{\perp}^{<}}$ diverges at 
the critical point with exponent $\nu_{\perp}^{<} \approx 2$. 

\section{Time-dependent behavior}
\label{dyn}

Starting from an initial 
state in which particles are Poisson distributed, we study the temporal 
decay of the activity at and below the critical point. 
The activity $S(t)$ at time $t$ obeys the following equation
\be
\frac{\partial S(t)}{\partial t} =\frac{1}{2 d} \sum_{\delta} \left[
\sum_{m=2} P_{k,k+\delta}(1,m,t)-\sum_{m=1}
P_{k,k+\delta}(2,m,t) 
-P_{k,k+\delta}(2,0,t) \right] \;. \l{time}
\ee
The first term on the right hand side (RHS) represents the gain in the 
activity when a particle hops out of an active site to 
an inactive site, whereas the last two loss terms 
correspond to a particle leaving an active site with two particles. 
At low densities, since the probability 
of having large mass is exponentially small, the first 
two terms on the RHS of Eq.(\ref{time}) can be ignored
and $S(t)$ can be approximated by $P(2,t)$. Denoting $m=0$ 
by $A$, $m=1$ by $\phi$ and
$m=2$ by $B$, the last term describes a two-species annihilation
reaction $A+B \tnd \phi$ where $A$ is a static and 
$B$ is a diffusing species. A similar mapping to reaction-diffusion system 
for the models in \cite{fesm1} involves more complicated 
reactions due to the creation of empty sites. 
In the following, we will denote the number of 
species $A$ and $B$ 
at time $t$ by $n_{A}(t)$ and $n_{B}(t)$ respectively.

At the critical density $\rho_{c}=1$, we expect 
that $n_{B}(0) \approx n_{A}(0)$ and in a finite domain of
length $\ell$, they differ by $\pm O(\ell^{d/2})$. Since
the species $B$ performs diffusive random walk, at large times in an
infinite system, we have \cite{AB0eq}
\be
S(\rho_{c}^{\infty},t) \sim \cases { t^{-d/4} & {, $d < 4$} \cr
        t^{-1} & {, $d \geq 4  \;\;,$}} \l{scp}
\ee
where $\rho_{c}^{\infty}$ refers to the critical density in the 
thermodynamic limit. 
For a finite system of size $L$, the activity $S$ at 
density $\rho_{c}^{\infty}$ is expected to be of the scaling form, 
\be
S(\rho_{c}^{\infty},t,L) \approx {t^{-\theta}}\; H (t/L^{z}) \;\;,\l{Scp}
\ee
where the scaling function $H(x)$ is a constant for $x \ll 1$ and
grows as $x^{\theta}$ for $x \gg 1$ \cite{fesm1}. In other words, the 
activity typically decays as a power law in time to a system size 
dependent constant $L^{-z \theta}$. As shown in 
Fig.(\ref{theta1d}), in one dimension, 
the activity $S$ is of the scaling form in Eq.(\ref{Scp}) 
with $z \approx 2$ and the exponent $\theta$ given by Eq.(\ref{scp}).  
However, the scaling function $H(x)$ \emph{decays} exponentially 
for $x \gg 1$ as seen by the linear decay of scaled activity 
on the semilog scale in Fig.(\ref{theta1d}). In two dimensions, our 
simulations support the scaling behavior Eq.(\ref{Scp}) with the scaling 
function decaying as a stretched exponential. The finite size scaling of the 
activity $S \propto L^{-z \theta}$ in the steady state does not hold since 
the exact 
expression for $S=(M-V)/(M-1)$ gives $\rho_{c}=1$ for \emph{any} $L$
(unlike other sandpiles where $\rho_{c}$ is expected to be size-dependent).

For $\rho \ll \rho_{c}$, we expect $n_{B}(0) \ll n_{A}(0)$ so that
$n_{A}$ does not change appreciably due to the annihilation reaction and
we are led to consider the problem of diffusing $B$ species in the presence
of static traps $A$. In this case, at large times,  
the density $S(t)$ decays as a stretched exponential \cite{traps},
\be
S(t) \sim \mbox{exp}\left[-(t/t_{d})^{\alpha} \right]
\;\;\;,\;\;\; \mbox{for} 
\;\;  \rho < \rho_{c}  \l{sdead} \;,
\ee
where $\alpha=d/(d+2)$ and $t_{d} \sim (1-\rho)^{-2/d}$. 
This slow decay arises due to large trap-free
regions which are rare but enhance survival significantly. The inset of 
Fig.(\ref{theta1d}) shows the decay of $S(t)$ in one dimension in 
accordance with Eq.(\ref{sdead}). This behavior of 
slow relaxation to the steady state is similar 
to that observed numerically in various stochastic FESMs in \cite{sexpo}.


We next consider the effect of a perturbation
in the steady state by considering the correlation function 
$G(\vec{r},t)=\langle \eta(0,0) \; \eta(\vec{r},t) \rangle$ 
where $\eta(\vec{r},t)$ is one if the site $\vec{r}$ is active at time $t$ 
and zero otherwise. We perturb the steady state by constructing an initial 
condition with a single mobile particle placed at the origin. 
At the critical point, this particle executes a random walk and the 
correlation function $G(\vec{r},t)$ is equal to the probability that the 
walker is at $(\vec{r},t)$ so that 
\be
G(\rho_{c},r,t) \approx 
\frac{1}{(2 \pi t)^{d/2}} \mbox{exp}({-r^{2}/2 t}) \;\;.
\ee

In the absorbing phase, there is a finite density $1-\rho$ of the 
vacant sites and the typical distance $r_{0}$ between them scales as 
$(1-\rho)^{-1/d}$. Therefore, the single mobile particle at the origin 
diffuses for $t \ll r_{0}^{2}$ so that the correlation function 
$G(\vec{r},t) \sim  e^{-r^{2}/2 t}, r < r_{0}$ 
increases with time. However, for 
$t \gg r_{0}^{2}$, the mobile particle gets trapped and $G(\vec{r},t)$ decays
as in Eq.(\ref{sdead}). Usually, the auto-correlation
function $G(0,t)$ is expected to decay 
exponentially, i.e. $G(0,t) \sim e^{-t/\tau}$ where 
$\tau \sim (\rho_{c}-\rho)^{-\nu_{||}}$ and the exponent $\nu_{||}$ 
obeys the scaling relation $\nu_{||}= \nu_{\perp} z$ \cite{abs}. The 
stretched exponential behavior of $G(0,t)$ in our model implies 
that $\nu_{||}^{<}$ is infinite. Alternatively, one can define this exponent 
via $t_{d} \sim (\rho_{c}-\rho)^{-\nu_{||}^{<}}$ with $\nu_{||}^{<}=2/d$. 
Since, in our model, $z \approx 2$ 
and the correlation length exponent $\nu_{\perp}^{<} \approx 2$ in one 
dimension, either definition of $\nu_{||}^{<}$ is inconsistent with the above 
scaling relation and we conclude that it does not hold in the 
absorbing phase. 


\section{Conclusion}
\l{concl}

We introduced a sandpile model whose simplicity 
allowed us to determine various static and dynamic exponents exactly. 
These exponents differ from those of the C-DP class for which, in one dimension, 
$\beta \approx 0.29$, $\nu_{\bot}^{>} \approx 1.33$, 
$z \approx 1.55$ and $\theta \approx 0.14$ \cite{cdp2}. 
We considered the correlation lengths 
$\xi^{>}$ and $\xi^{<}$ defined above and below the critical density 
respectively and found that while $\xi^{>}$ is zero, $\xi^{<}$ diverges as 
$\rho \tnd \rho_{c}-$. However, a further detailed study of the nontrivial 
correlations present in the 
inactive phase would be interesting. We also studied the temporal properties 
of this system by relating it to a well studied 
reaction-diffusion system via an argument similar to that used in 
\cite{krug} to study avalanche size distribution in a sandpile model. 

One can also consider the biased version of the above model in
which a particle moves preferentially to a nearest neighbor. The
steady state of this model is the same as that for the symmetric case 
discussed above. For the asymmetric case, using the known results for 
the two-species annihilation
problem with $B$ species drifting with a nonzero speed,
we expect that at the critical point, $S(t)$ decays as a
power law with an exponent $d/2$ for $d \leq 2$ and exponentially in  
the subcritical regime \cite{btraps}. Our numerical results are
consistent with the argument.

Acknowledgment: The author thanks D. Dhar for valuable 
discussions and for pointing out an error in an earlier version and J. Krug 
for comments on the manuscript. This work has been supported by DFG within 
SFB/TR 12 {\it Symmetries and Universality in Mesoscopic Systems}.


\begin{figure}
\begin{center}
\includegraphics[scale=0.35,angle=270]{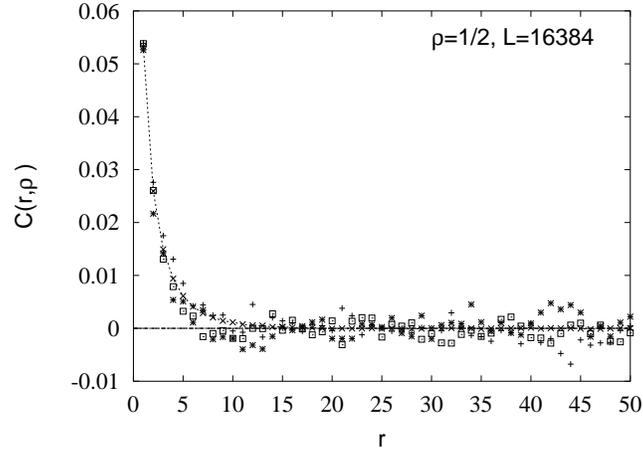}
\caption{Plot of $C(r,\rho)$ vs. $r$ for $\rho < \rho_{c}$ in one dimension 
with Poisson distributed initial mass. The data shown with points 
corresponds to different initial seeds for the random number generator. 
The data averaged over $1000$ initial conditions is shown with broken line.}
\l{ic}
\end{center}
\end{figure}

\begin{figure}
\begin{center}
\includegraphics[scale=0.35,angle=270]{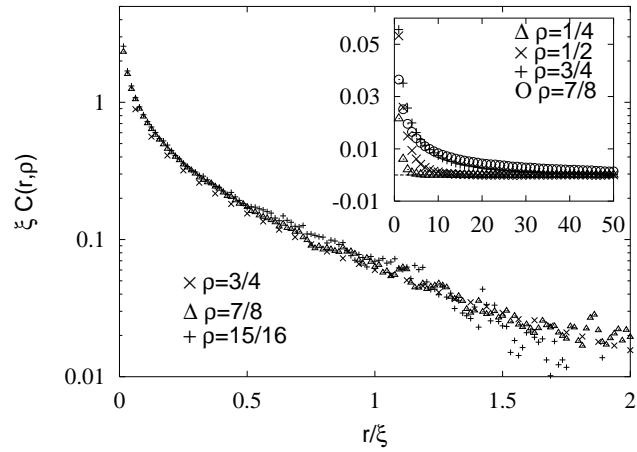}
\caption{
Inset: Plot of $C(r,\rho)$ vs. $r$ in one dimension for $\rho < \rho_{c}$ 
with $L=16384$ to show that the product measure does not hold in the 
inactive phase. Main: Data collapse for the scaled correlation 
function $\xi \; C(r,\rho)$ vs. $r/ \xi$ with $\xi=(1-\rho)^{-2}$.}
\l{xi}
\end{center}
\end{figure}

\begin{figure}
\begin{center}
\includegraphics[scale=0.35,angle=270]{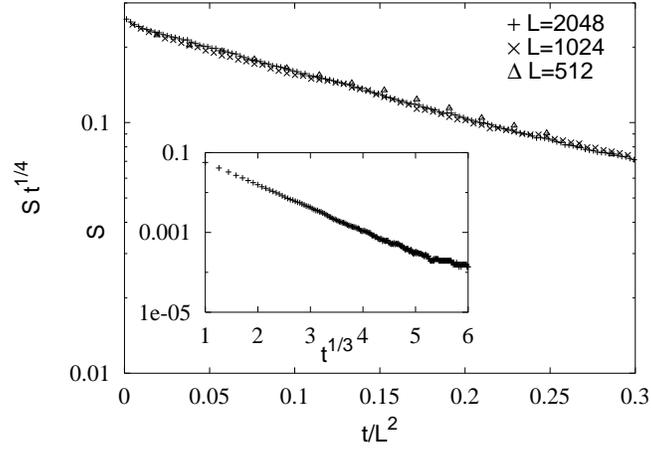}
\caption{Data collapse for the scaled activity  
$t^{1/4} S(t,L)$ vs. $t/ L^{2}$ at $\rho=\rho_{c}$ in $1d$ in accordance 
with Eq.(\ref{Scp}). The inset shows the temporal decay of $S(t)$ for 
$\rho < \rho_{c}$ as in Eq.(\ref{sdead}).}
\l{theta1d}
\end{center}
\end{figure}


\end{document}